\input harvmac

\def \p {\phi}
\def \ha {\half}
\def \ov {\over}

\def \a {\alpha}
\def \lr { \lref}
\def\ep{\epsilon}

\def \r {\rho}
 \def\l
{\lambda}\def \G {\Gamma}

\def\g {\gamma}

\def \I {{\cal I}}

\def \L {\Lambda}

\def \tr {{ \rm tr  }}
\def \k {\kappa}
\def \we {\wedge}
\def \cd {\cdot}
 \def \J {{\cal J}}
 \def \k {\kappa}
 \def \rf {\refs}
 \def \W {{\cal W}}

\def \k {\kappa}
\def \lr { \lref}
\def \C {{\cal C}}
\def \S  {{\cal S}}
\def \N {{\cal N}}

\def \adsss {$AdS_7 \times S^4$\ } 
\gdef \jnl#1, #2, #3, 1#4#5#6{ { #1~}{ #2} (1#4#5#6) #3}

\baselineskip8pt
\Title{\vbox
{\baselineskip 6pt{\hbox{OHSTPY-HEP-T-00-006  }}{\hbox
{   }}{\hbox{hep-th/0005072}} {\hbox{
   }}} }
{\vbox{\centerline {$R^4$ terms in $11$ dimensions }
\medskip
 \centerline {and   conformal anomaly of
 (2,0) theory }
 }}
\vskip -20 true pt

\medskip
\centerline{   A.A. Tseytlin\footnote{$^{\star}$}{\baselineskip8pt
e-mail address: tseytlin@mps.ohio-state.edu}\footnote{$^{\dagger}$}{\baselineskip8pt
Also at   Blackett Laboratory, Imperial 
College, London and  Lebedev  Physics
Institute, Moscow.} }

\smallskip\smallskip
\centerline {\it  Department of Physics }
\smallskip
\centerline {\it  The Ohio State University}\smallskip
\centerline {\it  Columbus, OH 43210-1106, USA}

\bigskip\bigskip
\centerline {\bf Abstract}
\medskip
\baselineskip10pt
\noindent
\medskip
Using AdS$_7$/CFT$_6$ correspondence
we compute a subleading $O(N)$ term in  the scale  anomaly
of  (2,0) theory describing  $N$ coincident M5 branes.
While the leading $O(N^3)$  contribution 
 to the  anomaly is determined by the value of 
the supergravity action, the $O(N)$ contribution  
comes  from  a particular $R^4$ term 
(8-d Euler density invariant)  in the 11-dimensional
effective action. 
 This $R^4$ term is argued 
to be part of the same 
superinvariant as the  P-odd   
$\C_3 R^4$ term  known to produce 
$O(N)$ contribution to the 
R-symmetry anomaly  of (2,0) theory.
The known  results for R-anomaly suggest that the 
total scale  anomaly  extrapolated to N=1 should 
be the same as the anomaly of a single free
(2,0) tensor multiplet. A proposed  explanation of 
 this agreement is  that the  coefficient  $4N^3$  in the anomaly 
(which was found previously to be also 
the ratio of the 2-point and 3-point  graviton correlators
in the (2,0) theory and in the free tensor multiplet theory)
is shifted to $4N^3 -3N$.

\Date {May  2000}

\noblackbox
\baselineskip 16pt plus 2pt minus 2pt

\lr \tse {A.A.~Tseytlin,
``Heterotic - type I superstring duality and low-energy
effective actions,''
Nucl.\ Phys.\  {\bf B467}, 383 (1996)
[hep-th/9512081].
 } 

\lr \gw {   D.J.~Gross and E.~Witten,
``Superstring Modifications Of Einstein's Equations,''
Nucl.\ Phys.\  {\bf B277}, 1 (1986).
}

\lr\gs{  M.B.~Green and J.H.~Schwarz,
``Supersymmetrical Dual String Theory. 2. Vertices And Trees,''
Nucl.\ Phys.\  {\bf B198}, 252 (1982).
 }

\lr\sak { N.~Sakai and Y.~Tanii,
``One Loop Amplitudes And Effective Action In Superstring
Theories,''
Nucl.\ Phys.\  {\bf B287}, 457 (1987).
 }

\lr\kp  {E.~Kiritsis and B.~Pioline,
``On $R^4 $  threshold corrections in type IIB string theory and
(p,q) string  instantons,''
Nucl.\ Phys.\  {\bf B508}, 509 (1997)
[hep-th/9707018].
}

\lr\ant  {
I.~Antoniadis, S.~Ferrara, R.~Minasian and K.S.~Narain,
``$R^4$ couplings in M- and type II theories on Calabi-Yau
spaces,''
Nucl.\ Phys.\  {\bf B507}, 571 (1997)
[hep-th/9707013].
}

\lr\vw {C.~Vafa and E.~Witten,
``A One loop test of string duality,''
Nucl.\ Phys.\  {\bf B447}, 261 (1995)
[hep-th/9505053].
}

\lr\ler { W.~Lerche, B.E.~Nilsson and A.N.~Schellekens,
``Heterotic String Loop Calculation Of The Anomaly Cancelling
Term,''
Nucl.\ Phys.\  {\bf B289}, 609 (1987);
  W.~Lerche, B.E.~Nilsson, A.N.~Schellekens and
N.P.~Warner,
``Anomaly Cancelling Terms From The Elliptic Genus,''
Nucl.\ Phys.\  {\bf B299}, 91 (1988).
   }

\lr\duf {M.J.~Duff, J.T.~Liu and R.~Minasian,
``Eleven-dimensional origin of string / string duality: A
one-loop test,''
Nucl.\ Phys.\  {\bf B452}, 261 (1995)
[hep-th/9506126].
}

\lr \myers{ 
R.~Myers,
``Superstring Gravity And Black Holes,''
Nucl.\ Phys.\  {\bf B289}, 701 (1987).
}

\lr\ren  { 
R.~Kallosh,
``Strings And Superspace,''
Phys.\ Scripta {\bf T15}, 118 (1987).
B.~E.~Nilsson and A.~K.~Tollsten,
``Supersymmetrization Of $\zeta(3) R^4$ In
Superstring Theories,''
Phys.\ Lett.\  {\bf B181}, 63 (1986);
}

\lr\sul {  M.~de Roo, H.~Suelmann and A.~Wiedemann,
``Supersymmetric $R^4$ actions in ten-dimensions,''
Phys.\ Lett.\  {\bf B280}, 39 (1992);
``The Supersymmetric effective action of the heterotic string
in ten-dimensions,''
Nucl.\ Phys.\  {\bf B405}, 326 (1993)
[hep-th/9210099];
J.~H.~Suelmann,
``Supersymmetry and string effective actions,''
Ph.D. thesis,  Groningen, 1994.
 }

\lr\sg {M.B.~Green and S.~Sethi,
``Supersymmetry constraints on type IIB supergravity,''
Phys.\ Rev.\  {\bf D59}, 046006 (1999)
[hep-th/9808061].
}

\lr\mt { R.R.~Metsaev and A.A.~Tseytlin,
``On Loop Corrections To String Theory Effective Actions,''
Nucl.\ Phys.\  {\bf B298}, 109 (1988).
}

\lr\gree{ M.B.~Green, M.~Gutperle and P.~Vanhove,
``One loop in eleven dimensions,''
Phys.\ Lett.\  {\bf B409}, 177 (1997)
[hep-th/9706175].
}

\lr\TF { A.A.~Tseytlin,
``Ambiguity In The Effective Action In String Theories,''
Phys.\ Lett.\  {\bf B176}, 92 (1986).
}

\lr\gkt { S.S.~Gubser, I.R.~Klebanov and A.A.~Tseytlin,
``Coupling constant dependence in the thermodynamics of N = 4 
 supersymmetric Yang-Mills theory,''
Nucl.\ Phys.\ {\bf B534}, 202 (1998), 
hep-th/9805156.
}

\lr\bg { T.~Banks and M.B.~Green,
``Non-perturbative effects in $AdS_5 \times  S^5$ string theory and
d = 4 SUSY  Yang-Mills,''
JHEP {\bf 9805}, 002 (1998)
[hep-th/9804170].
}

\lr \fer {E.~Cremmer and S.~Ferrara,
``Formulation Of Eleven-Dimensional Supergravity In
Superspace,''
Phys.\ Lett.\  {\bf B91}, 61 (1980);
L.~Brink and P.~Howe,
``Eleven-Dimensional Supergravity On The Mass - Shell In
Superspace,''
Phys.\ Lett.\  {\bf B91}, 384 (1980).
}

\lr \kal { 
R.~Kallosh and A.~Rajaraman,
``Vacua of M-theory and string theory,''
Phys.\ Rev.\  {\bf D58}, 125003 (1998)
[hep-th/9805041].
}

\lr \KT { I.R.~Klebanov and A.A.~Tseytlin,
``Entropy of Near-Extremal Black p-branes,''
Nucl.\ Phys.\ {\bf B475}, 164 (1996), 
hep-th/9604089.
}

\lr \odin{S.~Nojiri and S.~D.~Odintsov,
``Weyl anomaly from Weyl gravity,''
hep-th/9910113;
 ``On the conformal anomaly from higher derivative gravity in
 AdS/CFT  correspondence,''
hep-th/9903033.
 }
   
\lr \FT{E.S.~Fradkin and A.A.~Tseytlin,
``Conformal Anomaly In Weyl Theory And Anomaly Free
Superconformal
Theories,''
Phys.\ Lett.\  {\bf B134}, 187 (1984).
   }

\lr \bon  { L.~Bonora, P.~Pasti and M.~Bregola,
``Weyl Cocycles,''
Class.\ Quant.\ Grav.\ {\bf 3}, 635 (1986).
  }
\lr \schw   { S.~Deser and A.~Schwimmer,
``Geometric classification of conformal anomalies in arbitrary
dimensions,
Phys.\ Lett.\  {\bf B309}, 279 (1993),
hep-th/9302047.
 } 

\lr \bft {  F.~Bastianelli, S.~Frolov and A.A.~Tseytlin,
``Conformal anomaly of (2,0) tensor multiplet in six
dimensions and  AdS/CFT correspondence,''
 JHEP  {\bf  02}, 013 (2000), 
hep-th/0001041.} 

\lr \blau  { M.~Blau, K.~S.~Narain and E.~Gava,
``On subleading contributions to the AdS/CFT trace anomaly,''
JHEP {\bf 9909}, 018 (1999)
[hep-th/9904179].   } 

\lr\imbim { C.~Imbimbo, A.~Schwimmer, S.~Theisen and
S.~Yankelowicz,
``Diffeomorphisms and Holographic Anomalies,'' 
 Class.\ Quant.\ Grav.\  {\bf 17}, 1129 (2000), 
 hep-th/9910267.
 } 
 
\lr\witte { E.~Witten,
``Five-brane effective action in M-theory,''
J.\ Geom.\ Phys.\  {\bf 22}, 103 (1997), 
hep-th/9610234;
}

\lr\ft {E.S.~Fradkin and A.A.~Tseytlin,
``Quantum Properties Of Higher Dimensional And Dimensionally
Reduced Supersymmetric Theories,''
Nucl.\ Phys.\  {\bf B227}, 252 (1983).
``Present State Of Quantum Supergravity,''
 In {\it Proc. of Third Seminar on Quantum Gravity}, 
Oct. 23-25 1984, Moscow, M.A. Markov et al eds., 
 p. 303 (World Scientific, 1985)
}

\lr\HS {  M.~Henningson and K.~Skenderis,
``The holographic Weyl anomaly,''
JHEP {\bf 07}, 023 (1998),
hep-th/9806087.}

\lr \harv {J.A.~Harvey, R.~Minasian and G.~Moore,
``Non-abelian tensor-multiplet anomalies,''
JHEP {\bf 09}, 004 (1998),
hep-th/9808060. } 

\lr  \gv  {M.B.~Green and P.~Vanhove,
``D-instantons, strings and M-theory,''
Phys.\ Lett.\  {\bf B408}, 122 (1997)
[hep-th/9704145].
 }

\lr \keh   {  D.~Anselmi and A.~Kehagias,
``Subleading corrections and central charges in the AdS/CFT 
correspondence,''
Phys.\ Lett.\  {\bf B455}, 155 (1999)
[hep-th/9812092].
} 

\lr \rs  { J.G.~Russo and A.A.~Tseytlin,
``One-loop four graviton amplitude in eleven dimensional
supergravity,''
Nucl.\ Phys.\  {\bf B578}, 139 (2000),
hep-th/9707134.
} 

\lr \basf{ 
F.~Bastianelli, S.~Frolov and A.A.~Tseytlin,
``Three-point correlators of stress tensors in
maximally-supersymmetric  
conformal theories in $d = 3$ and $d = 6$,''
Nucl.\ Phys.\  {\bf B508}, 245 (1997),
hep-th/9911135.
}

\lr\alv{L.~Alvarez-Gaume and E.~Witten,
``Gravitational Anomalies,''
Nucl.\ Phys.\ {\bf B234}, 269 (1984).
}

\lr\freedm { D.~Anselmi, D.Z.~Freedman, M.T.~Grisaru and
A.A.~Johansen,
``Nonperturbative formulas for central functions of
supersymmetric gauge  theories,''
Nucl.\ Phys.\  {\bf B526}, 543 (1998)
[hep-th/9708042].
}

\lr\GK { S.S.~Gubser and I.R.~Klebanov,
``Absorption by branes and Schwinger terms in the world volume
theory,''
Phys.\ Lett.\  {\bf B413}, 41 (1997)
[hep-th/9708005].
}

\lr \grisa { M.T.~Grisaru, A.E.~van de Ven and D.~Zanon,
``Four Loop Beta Function For The N=1 And N=2 Supersymmetric
Nonlinear Sigma Model In Two-Dimensions,''
Phys.\ Lett.\  {\bf B173}, 423 (1986);
``Four Loop Divergences For The N=1 Supersymmetric Nonlinear
Sigma Model In Two-Dimensions,''
Nucl.\ Phys.\  {\bf B277}, 409 (1986).
}

\lr \zan{M.T.~Grisaru and D.~Zanon,
``Sigma Model Superstring Corrections To The Einstein Hilbert
Action,''
Phys.\ Lett.\  {\bf B177}, 347 (1986);
M.D.~Freeman, C.N.~Pope, M.F.~Sohnius and K.S.~Stelle,
``Higher Order Sigma Model Counterterms And The Effective Action
For Superstrings,''
Phys.\ Lett.\  {\bf B178}, 199 (1986);
Q.~Park and D.~Zanon,
``More On Sigma Model Beta Functions And Low-Energy Effective
Actions,''
Phys.\ Rev.\  {\bf D35}, 4038 (1987).
}

\lr \freed {D.~Freed, J.A.~Harvey, R.~Minasian and
G.~Moore,
``Gravitational anomaly cancellation for M-theory
fivebranes,''
Adv.\ Theor.\ Math.\ Phys.\  {\bf 2}, 601 (1998)
[hep-th/9803205].
}

\lr \howe { P.S.~Howe, G.~Sierra and P.K.~Townsend,
``Supersymmetry In Six-Dimensions,''
Nucl.\ Phys.\  {\bf B221}, 331 (1983).
  }

\lr \berg{E.A.~Bergshoeff and M.~de Roo,
``The Quartic Effective Action Of The Heterotic String And
Supersymmetry,''
Nucl.\ Phys.\  {\bf B328}, 439 (1989).
}

\lr \green{  M.B.~Green,
``Interconnections between type II superstrings, M theory and N =
4  Yang-Mills,''
hep-th/9903124.}

\lr \how{ 
P.S.~Howe and P.C.~West,
``The Complete N=2, D = 10 Supergravity,''
Nucl.\ Phys.\  {\bf B238}, 181 (1984).
}

\lr \kep{A.~Kehagias and H.~Partouche,
``The exact quartic effective action for the type IIB
superstring,''
Phys.\ Lett.\  {\bf B422}, 109 (1998)
[hep-th/9710023].
}

\lr \zumino{B.~Zumino,
``Gravity Theories In More Than Four-Dimensions,''
Phys.\ Rept.\  {\bf 137}, 109 (1986).
}

\lr\pilch{K.~Pilch, P.~van Nieuwenhuizen and P.K.~Townsend,
``Compactification Of D = 11 Supergravity On $S^4$ (Or 11 = 7 + 4,
Too),''
Nucl.\ Phys.\  {\bf B242}, 377 (1984).
}

\lr\mald{J.~Maldacena,
``The large N limit of superconformal field theories and
supergravity,''
Adv.\ Theor.\ Math.\ Phys.\  {\bf 2}, 231 (1998)
[hep-th/9711200].
}

\lr \LT {  H.~Liu and A.A.~Tseytlin,
``D = 4 super Yang-Mills, D = 5 gauged supergravity, and D = 4
conformal  
supergravity,'' Nucl.\ Phys.\ {\bf B533}, 88 (1998), 
hep-th/9804083.
}

\lr \GKT {  S.S.~Gubser, I.R.~Klebanov and A.A.~Tseytlin,
``String theory and classical absorption by three-branes,''
Nucl.\ Phys.\ {\bf B499}, 217 (1997), 
hep-th/9703040. 
}

\lr \pet  { R.~Manvelian and A.C.~Petkou,
``A note on R-currents and trace anomalies in the (2,0) tensor
multiplet  in d = 6 AdS/CFT correspondence,''
hep-th/0003017.
 }

\lr  \BFT{    F.~Bastianelli, S.~Frolov and A.A.~Tseytlin,
``Three-point correlators of stress tensors in
maximally-supersymmetric  
conformal theories in $d = 3$ and $d = 6$,''
hep-th/9911135.
}

\lr \AF { G.~Arutyunov and S.~Frolov,
``Three-point Green function of the stress-energy tensor in the
AdS/CFT  
correspondence,'' Phys.\ Rev.\ {\bf D60}, 026004 (1999), 
hep-th/9901121.
} 

\lr \IK {I.R.~Klebanov,
``World-volume approach to absorption by non-dilatonic
branes,''
Nucl.\ Phys.\  {\bf B496}, 231 (1997)
[hep-th/9702076].
 }

\lr\fef{C. Fefferman and C. Graham, 
``Conformal invariants", 
Asterisque, hors serie, 95 (1985);
T.~Parker and S.~Rosenberg,
``Invariants of conformal Laplacians", 
 J.\ Diff.\ Geom.\ {\bf 25,} 199 (1987);
J.~Erdmenger,
``Conformally covariant differential operators: Properties and 
applications,''
Class.\ Quant.\ Grav.\  {\bf 14}, 2061 (1997), 
hep-th/9704108.
}
\lr\dess{
S.~Deser and D.~Seminara,
``Tree amplitudes and two-loop counterterms in D = 11
supergravity,''
hep-th/0002241.
}

\newsec{Introduction}

Two known  maximally (2,0) supersymmetric  conformal field 
theories in 6 dimensions are the free tensor multiplet
theory describing  low energy dynamics of
a single M5 brane, and  still largely mysterious  
interacting (2,0) conformal theory 
describing $N$ coincident M5 branes.  
A way to study the latter theory is provided 
by its  conjectured  duality \mald\  to 
M-theory (or, for large $N$, 11-d supergravity corrected by
higher derivative terms) 
on \adsss background. 

Comparison of  the  2-point   and 
3-point correlators of the 
stress tensor  of (2,0) theory 
as predicted by the  $AdS_7 \times S^4$  supergravity 
\rf{\LT,\AF} 
 to those  in the free tensor multiplet theory 
shows   \rf{\BFT,\GKT,\GK}\  that they differ 
only  by the
 overall coefficient
  $4N^3$.\foot{The same is true also
   for the correlators of R-symmetry
 currents \pet.}
The remarkable coefficient $4N^3$ 
 was  originally  found in \rf{\GKT} 
in the comparison 
of the  M5 brane  world volume  theory  and the $D=11$ 
supergravity expressions  for the 
absorption cross-sections of longitudinally polarized 
gravitons by  $N$ coincident  M5 branes.
The same coefficient  $4N^3$ 
appears also as the ratio 
of the   scale anomalies 
 (or Weyl-invariant parts of 
 conformal anomalies)  of the interacting (2,0) theory \HS\
 and  free theory of a single  tensor multiplet \bft.
   
The reason why the coefficient $4N^3$ was 
puzzling in \GKT\  was  analogy with the $d=4$
case: a similar comparison
of the gravitational and  world-volume  absorption
cross-sections in the case of  D3-branes \rf{\IK,\GKT}
led to the ratio $N^2$, which  is equal to 1 for $N=1$.
This agreement in the $d=4$  case was later understood \GK\ as
being a consequence of nonrenormalization of the conformal
anomaly and thus of the 2-point stress tensor correlator 
in $\N=4$ SYM theory.  The  analogy  between the $d=4$ and
$d=6$ cases 
should  not, of course,  be taken too seriously,  
given 
 that the (2,0)  theory  should  have a  different 
 structure than  SYM theory,
 being  an interacting conformal fixed point without
 a free coupling parameter.
 
 Still, one may expect that anomalies and 2- and 3-point
 correlators of  currents  of the (2,0) theory may 
 have special ``protected" form,  with simple dependence on
 $N$, allowing one to interpolate between $N \gg 1 $ and $N=1$ 
 cases. 

This was, in fact, observed  for  the
R-symmetry anomaly of the (2,0) theory \harv:
the anomaly of the (2,0) theory obtained from 
the 11-d action containing the standard 
supergravity  term  plus  a higher-derivative 
$\C_3 R^4$ 
term \duf\ is given by the sum of the 
leading supergravity $O(N^3)$ and subleading  $O(N)$ terms, 
and  for $N=1$ is equal to the R-symmetry anomaly 
corresponding to the single tensor multiplet \rf{\alv,\witte}.

Since the conformal and R-symmetry anomalies of 
the  (2,0) theory should belong to the same $d=6$ 
supermultiplet
\rf{\howe,\harv}, 
 one should then 
expect to find  a similar $O(N)$  correction 
to the  $O(N^3)$ supergravity  contribution  \HS\ 
 to the (2,0)  conformal anomaly. This  $O(N)$ correction   
 should originate from a    higher-derivative 
$R^4$  term in the 11-d action which should be a part of the same
superinvariant as  $\C_3R^4$ term (just 
like the second-derivative supergravity terms $R$ and
  $\C_3 F_4 F_4$ are).
 
Our aim below is to  discuss  a mechanism of how this
 may happen.
 We shall argue that the 
 11-d  action contains  a  particular $R^4$  term,  
  which, upon compactification on $S^4$, leads
 to  a special combination of  $R^3$  terms in the effective 7-d action.
 These  $R^3$ corrections  produce    extra $O(N)$ 
 terms in the conformal anomaly of the boundary (2,0) 
 conformal theory. As a result, 
  the coefficient $4N^3$ 
 in the ratio of the (2,0) theory and  tensor multiplet
  scale anomalies  may be 
 shifted to $4N^3-3N$.  Since the 
  latter is  equal to 1 for $N=1$, 
  this would be a resolution of  the ``$4N^3$" puzzle. 
 
Since  this  conclusion is  sensitive to numerical values
of  coefficients in the 11-d low energy effective action
we shall start with a  critical 
 review  of what is known 
 about  the structure of $R^4$ terms in type
IIA string theory  in 10-d and their counterparts in  M-theory. 
While the type IIB theory effective action
 contains the same $J_0 \sim R^4$ 
invariant  at the tree and one-loop levels, 
the  one-loop term in type IIA theory  
is a combination of two  different $R^4$ structures.
We shall argue that they  should be  organized into two 
different $\N=2A$ superinvariants --  $J_0$ and $\I_2$ 
(containing  P-odd \  $B_2 \tr R^4$ term) 
 in a way 
different than it  was previously suggested (Section 2).
The  corresponding  tow  $D=10$  superinvariants ``lifted"
 to $D=11$  represent   the 
leading $R^4$ corrections 
to the  11-d supergravity action (Section 3).

These terms should be supplemented by proper 
 $F_4=d\C_3$ dependent 
terms  as required by supersymmetry and  chosen in a specific 
``on-shell" scheme  not to   modify the \adsss solution
of the   $D=11$  supergravity. Assuming that, 
in  Section 4 we  discuss   higher
derivative corrections to  the 
 7-d action  of $S^4$ compactified theory 
 which  follow from  
 the presence of the $R^4$ terms
 in $D=11$ action. 
 In Section 5 we compute the corresponding $O(N)$ 
 contributions to the scale anomaly  of the (2,0) 
 theory  using  the method of \HS, 
 and  draw analogy  between  the total $O(N^3) + O(N)$ 
 result  and  the expression 
 for the R-symmetry  anomaly  found in \harv.

\newsec{$R^4$ terms in 10  dimensions}

Let us start with a review of the structure of the $R^4$ 
terms in the effective actions of type IIA superstring 
in 10 dimensions and the corresponding terms in 
M-theory  effective action in 11 dimensions,   
 paying  special  attention to explicit values of 
numerical coefficients.


The relevant terms in the tree + one loop  type IIA string theory
 effective
action can be written in the form
$$
S= S_0 + S_1 \ , 
$$
\eqn\tree{
S_0 =
 {1\over 2\kappa^2_{10}} \int d^{10} x \sqrt {-G} \  e^{-2\p} \ 
\ \bigg( R  - { 1 \ov 2\cdot 3!} H^2_3 + ...  +
  b_0 \a'^3 J_0 \bigg) \ , \ \ 
}
\eqn\loop{
S_1 =  { 1 \ov 2 \pi \a'}  \int d^{10} x \sqrt {-G} \ 
\  L_1 \ , \ \ \ \ \ \  \ \ \   L_1 =    b_1  \J_0   + b_2  K  
 \ , 
}
where  $H_{mnk} = 3 \del_{[m} B_{nk]}$ and\foot{We use Minkowski 
notation for
the metric  and $\ep$ tensor, so that $\ep_{10} \ep_{10} =-10!$, 
and upon reduction to 8 spatial 
 dimensions $ \ep_{mn...} \ep_{mn...} \to  - 2 \ep_8\ep_8$.
 For other notation see also \tse.}
\eqn\jo{
 J_0 =\J_1 + \J_2 \equiv   t_8\cdot t_8 RRRR +  {1\ov 4 \cdot 2!} \ep_{10}\cdot \ep_{10} RRRR\ ,
 }
 \eqn\joq{
\J_0 = \J_1 -  \J_2\equiv t_8\cdot t_8 RRRR -  {1\ov 4\cdot 2!} \ep_{10}\cdot \ep_{10} RRRR\ ,
 }
 \eqn\joqa{
 K =  \ep_{10} B_2  [ \tr R^4  - { 1 \ov 4} (\tr R^2)^2] \ . }
 In the notation we are using the numerical coefficients are 
\eqn\tyr{ 2 \k^2_{10} = (2\pi)^7 g_s^2 \a'^4  \ , }
\eqn\nuu{
 b_0 = { 1  \ov 3 \cdot 2^{11} } \zeta(3)   \ , \ \ \ \
 b_1 = { 1 \ov (2\pi)^4 \cd 3^2 \cd 2^{13}} \ , \ \ \ \ \ \ 
 b_2 = -12 b_1 = 
- { 1 \ov (2\pi)^4 \cd 3 \cd 2^{11} } \  . }
The  tree and one-loop coefficients of the 
well-known 
 $\J_1= t_8\cdot t_8 RRRR$ term\foot{The more explicit form of this
 term is 
 $ \J_1  =  24 t_8 [ \tr R^4  - { 1 \ov 4} (\tr
R^2)^2]\ ,$ where  $R = (R^{ab}_{\ \ mn})$ and 
$ 
t_8 \tr R^4 \equiv  \tr \big(   16 R_{mn}R_{rn}R_{ml}R_{rl} +
8 R_{mn}R_{rn}R_{rl} R_{ml} 
-   4 R_{mn}R_{mn}R_{rl}R_{rl}
- 2 R_{mn}R_{rl}R_{mn}R_{rl}\big)\ .$}
can be  determined  from  the 4-graviton amplitude
\rf{\gs,\gw, \sak}.\foot{Note that the total coefficient 
of the  $t_8\cdot t_8 RRRR$ term in $S$  is thus 
$ - { 1 \ov (2\pi)^7 \cd 3 \cd 2^{11} \a' } 
({ \zeta (3) \ov g_s^2} + {\pi^2 \ov 3})$.
The relative   combination $  { \zeta (3) \ov g_s^2} 
+ {\pi^2 \ov 3}$
is the same as in  \sak\ (where $g^2=
 (2\k_{10})^2 (2\a')^{-4} = 16 \pi^7 g^2_s$)
 and in \gv, but our 
overall normalization of this term  is
different (by factor $2^5$ compared to \gv).}

The invariant
 $\J_2 = { 1\ov 4\cd 2!} \ep_{10}\cdot \ep_{10} RRRR$
which will play   important  role in  what follows 
is the $D=10$  
extension  of the integrand of the Euler invariant 
 in 8 dimensions
\eqn\gb{\J_2 = { 1 \ov 4} E_8\ , \ \ \ \ \ \ \ \ \ \ 
E_8 = { 1 \ov (D-8)!} \ep_D \ep_D  R^4  
= \pm 8!\delta^{n_1}_{[m_1}
\ldots \delta^{n_8}_{m_8]} R^{m_1 m_2}_{\ \ \ \ \ n_1 n_2} \ldots R^{m_7 m_8}_
{\ \ \ \ \  n_7 n_8}
\ , }
where $\pm$ correspond to the case of 
 Euclidean or Minkowski signature.\foot{
 The Euler number in 8 dimensions is 
 $\chi = { 1 \ov (4\pi)^4\cd  3 \cd 2^{7} } \int d^8 x \sqrt{
 g} \ E_8 $.} 
 
The  expansion of $E_8$ 
near flat space ($g_{mn} = \eta_{mn} + h_{mn}$) 
 starts with $h^5$ terms (see, e.g., \zumino), so that  its coefficient 
cannot be directly determined from  the on-shell 
4-graviton amplitude. 
The sigma-model approach implies \refs{\grisa,\zan}
that $E_8$ does appear in $S_0$, i.e.
that (up to  usual field redefinition ambiguities) 
the tree-level type II string  $R^4$ term  is  indeed 
proportional to $J_0$ \jo.

The  structure of the kinematic factor
$ (t_8 + { 1\ov 2} \ep_8)
(t_8 + { 1\ov 2}  \ep_8) $ in  the one-loop 
type IIA  4-point amplitude with transverse polarisations and
momenta
suggests \rf{\rs,\kp,\ant}  that the one-loop 
$R^4$ terms in $D=10$ type IIA theory 
should be proportional 
to the  opposite-sign combination $\J_0$ \joq\ 
of the  $\J_1$ and $\J_2$ terms,  and  
this assumption passes some  compactification tests \rf{\kp,\ant}.

The presence of the  P-odd one-loop  term $K$ \joqa\ 
can be established \vw\  following similar calculations 
of anomaly-related terms in the heterotic string \ler. 
Its coefficient $b_2$ can be fixed by considering 
compactification to 2 dimensions \vw, and its value is 
in agreement  with the coefficient required by 5-brane anomaly 
cancellation \duf\ (see also below).

The low-energy effective string  action  should 
 be supersymmetric.\foot{The string  S-matrix is invariant 
 under on-shell supersymmetry, so the leading-order corrections
 to effective action evaluated on the supergravity 
 equations of motion 
 should  be invariant under the standard supersymmetry 
 transformations. Since the $D=10$ supersymmetry algebra does not
 close off shell, the full  off-shell effective action should be
 invariant under 
 ``deformed" supersymmetry transformations (see, e.g., 
 \berg).}
 Remarkably, the coefficients in  \nuu\ 
 are indeed consistent with  what is known 
 about the structure of  possible  $R^4$ super-invariants.
 First, the $h^4$ term in $t_8t_8 R^4$ is 
 the bosonic part of the on-shell linearized 
  superspace invariant \ren \ (i.e. $\int d^{16} \theta
\   \Phi^4$, \ $\Phi=\phi + ..+ \theta^4
  R + ...$ written  in terms of $\N=1$ or $\N=2B$ \rf{\how,\green} 
   on-shell superspace  superfield $\Phi$).
 If one first  restricts consideration to  $\N=1,\ D=10$
supersymmetry only, then  one can use the classification
of  possible  bosonic $R^4$ 
 parts of  on-shell 
non-linear  $\N=1$ superinvariants
given   in \sul. A basis of the three
 independent $\N=1$ invariants \rf{\sul,\tse}
can be chosen as $J_0, X_1,X_2$ 
\eqn\inva{
J_0  =    t_8\cdot t_8 RRRR +  {1\ov 4 \cdot 2!} \ep_{10}\cdot
\ep_{10} RRRR\      \ ,  }
\eqn\yut{
X_1 = t_8 \tr R^4 - { {1\ov 4}} \ep_{10} B_2  \tr R^4 \ ,
   \ \ \ \ \ \ 
X_2 = t_8 \tr R^2 \tr R^2  - { {1\ov 4}} \ep_{10} B_2  \tr
R^2 \tr R^2
\ . }
One  may try to  combine these $\N=1$ invariants
to form  potential $\N=2A$ 
superinvariants.
Since 
$t_8t_8 R^4 = 24  t_8[  \tr R^4 - { 1 \ov 4}
 \tr R^2 \tr R^2] $,  one   may consider 
 two  candidate invariants which contain  combinations of 
 $\J_1$ \jo\ or $\J_2$ \joq\   with $\pm 6 K$ \joqa, i.e.   
$$ {\cal I}_1= 24 ( X_1 - { 1 \ov 4} X_2)=\J_1  - 6 K $$ 
 \eqn\comb{= \ 
   t_8\cdot t_8 RRRR  -  6 \ep_{10} B_2  [ \tr R^4  - { 1 \ov 4} (\tr
   R^2)^2]  \ , }
   or  
$$
{\cal I}_2 =   J_0  - 24 ( X_1 - { 1 \ov 4} X_2)
= \J_2 + 6 K
$$
 \eqn\combi{= 
{1\ov 4 \cdot 2!} \ep_{10}\cdot \ep_{10} RRRR  + 6 \ep_{10} B_2  [ \tr R^4  - { 1 \ov 4} (\tr
   R^2)^2]  \ , }
   \eqn\coi{\ \ \ \  {\cal I}_1 + {\cal I}_2 = J_0 \ .    } 
The 1-loop  term $L_1$  \loop\  
 with $b_2= - 12 b_1$ 
 can thus be  represented  as 
  a combination of {\it two} different
   $R^4$ superinvariants \rf{\rs,\kp}, i.e. as  
 \eqn\lop{ 
 L_1=  b_1  \J_0   + b_2  K
  = b_1 ( \J_1 - \J_2 - 12 K) 
  = b_1 ( -  J_0  + 2 {\cal I}_1 ) \ ,  }
  or as 
 \eqn\lopp{
 L_1=   b_1 ( J_0  -  2 {\cal I}_2 ) \ .  }  
  The  $J_0$-term
 should represent   a  separate $\N=2$
 invariant.\foot{In
\sul\ where  non-linear extensions of $\N=1$  on-shell
$R^4$ superinvariants were constructed 
 the transformation of the dilaton  prefactor
 was ignored. As a result,  one was not able to  make a
 distinction between  $J_0$ terms appearing
at  the  tree and 1-loop levels. It is natural 
to conjecture  that $f(\p) J_0$ terms should combine into 
an  $\N=2A$  superinvariant 
(invariant under deformed supersymmetry).  
For a discussion of 
 supersymmetry of $e^{-2\p} R + f(\p) J_0$ 
action in type  IIB  supergravity 
theory see \sg.}
A non-trivial question is which of  ${\cal I}_1$ and ${\cal I}_2$ 
 can be  actually 
extended to an  invariant of $\N=2A$ supersymmetry.\foot{Once 
the dilaton dependence of  $J_0$ terms
is taken into account, one will not be able to 
freely switch between   ${\cal I}_1$  and ${\cal I}_2$
using \coi.}

We would like to argue that
it is ${\cal I}_2$  and not ${\cal I}_1$ 
 that is the true $\N=2A$ 
superinvariant.
Namely,  it is the  Euler  term $\J_2={ 1 \ov 4} E_8$  
and {\it not} 
 $\J_1=t_8t_8 RRRR$ that is the ``superpartner"
of the  $B_2$-dependent term $K$ \joqa.
The  form of the 1-loop correction $L_1$ that admits a
super-extension is then \lopp\ and not \lop.
Then the tree + one-loop $J_0$ terms 
in the type IIA theory will be exactly the {\it same}
as in the type IIB theory, 
$ - { 1 \ov (2\pi)^4 \cd 3 \cd 2^{13}\a' } 
({ \zeta (3) \ov g_s^2} + {\pi^2 \ov 3}) J_0$, 
with the  type IIA theory action containing 
in addition one 
 extra one-loop
contribution \lopp\ proportional to the 
superinvariant ${\cal I}_2$.

 Indeed, the weak-field expansions
 of  both $E_8$ and $K$  start with 5-order 
 terms, and the corresponding 5-point amplitudes
 should be related by global supersymmetry. 
At the same time, it  is hard to imagine  how 
 the linearized on-shell ``$\W^4$" $\N=2$ superspace
 invariant corresponding to $h^4$ term in $t_8t_8 RRRR$ 
   may have a non-linear extension
 containing P-odd term $K$. 
 
A more serious  argument against $t_8t_8 RRRR$ being
 a ``superpartner"
of $ \ep_{10} B_2  [ \tr R^4  - { 1 \ov 4} (\tr
   R^2)^2]$ is the following.
The  $D=10$  type II supergravity is known to  contain
a  one-loop
quadratic $\L^2$   UV divergence proportional to $t_8t_8 RRRR$
(this can be seen \mt\   by taking the field theory limit,  
$\a'\to 0, \ \L=$fixed, 
 in the  one-loop 4-graviton amplitude, 
cf. \loop). 
 At the same time, 
   the Chern-Simons type terms like   $\ep_{10} B_2  R^4$ 
can not appear  in the {\it UV divergent}  part
of  one-loop effective action.\foot{Known examples
of induced CS terms  have finite coefficients 
and originate from IR effects (they appear 
from 1-loop contributions containing 
$1\ov \del^2$ massless poles, 
and  thus can be re-written 
in a  manifestly gauge invariant but nonlocal form).}
This can be proved  directly by  using the 
background field method:
all one-loop  UV divergent terms must
be  manifestly invariant  under 2-form gauge transformations and
as well as diffeomorphisms. Since, e.g., 
a proper time cutoff is expected to
preserve supersymmetry at the level
 of one-loop  UV divergences, 
one  concludes that   $\J_1$  and $K$
 can not be  parts of the
same superinvariant.

Similar argument can be given in the context 
of $D=11$ theory.  
The  $t_8t_8 RRRR$ term   appears \rf{\ft,\gv,\gree,\rs}
as a cubic UV  divergence (with a particular value of the 
UV  cutoff being  fixed by duality considerations \gree), 
but  $\ep_{11} \C_3 R^4$ term \duf\  can  have   only 
   a finite coefficient 
(with a non-perturbative dependence on
$\k_{11}$ on dimensional grounds). 
Thus
(contrary to some previous suggestions  in the literature,
 cf.  \rf{\gv,\rs,\kp,\keh}) 
these terms  
can not be related
by supersymmetry, and 
the superpartner of the $\ep_{11} \C_3 R^4$ term 
should  be the $D=11$ analog of  $\J_2={ 1 \ov 4} E_8$
(see section 3).

Before  turning to a detailed  discussion
of   the  $D=11$ terms, 
let us add few 
comments  about  the structure of the $D=10$  effective action
\tree,\loop.
In addition to the $R^4$ terms given explicitly
 in \jo\ and \joq,  it may contain also other
 Ricci tensor dependent terms 
 as well as  terms  depending on other fields (cf. \kep), 
 for example, terms involving two and more powers 
 of $H_3=dB_2$ (which were not included in the discussion 
 of super-invariants in \sul).
The well-known  field redefinition ambiguity \rf{\gw,\TF}
allows one to change the coefficients of ``on-shell" 
 terms.\foot{For example,  ignoring other fields, 
 one may use $R_{mn}=0$ to simplify 
the structure of $R^4$ invariants   as
the graviton legs in the  string amplitudes they correspond
to are on mass shell.}
In particular, the tree-level effective action \tree\ 
 may contain 
other $R_{mn}$ dependent terms in addition to 
the full curvature contractions present  in 
$J_0$  (see \rf{\zan,\myers,\sul}) 
\eqn\jjjk{
J_0  =  3 \cdot 2^8 ( R^{hmnk} R_{pmnq} R_{h}^{\ rsp}
 R^{q}_{\ rsk} 
          \ + \ \half  R^{hkmn} 
          R_{pqmn} R_h^{\ rsp} R^{q}_{\ rsk} )  + O(R_{mn})\ . }
The field redefinition ambiguity 
allows one to choose the action in a specific  ``scheme"
  where only the  Weyl tensor 
part of the curvature  appears in $J_0$, i.e. 
\eqn\jjj{
J_0\  \to\ \hat  J_0  \  =  3 \cdot 2^8 ( C^{hmnk} C_{pmnq} C_{h}^{\ rsp}
 C^{q}_{\ rsk} 
          \ + \ \half  C^{hkmn} 
          C_{pqmn} C_h^{\ rsp} C^{q}_{\ rsk} ) \ . }
That freedom of choice of a special  scheme is crucial, in 
particular, 
in order to  avoid corrections to certain highly symmetric
 leading-order solutions, both in 10 and in 11 dimensions 
 (see section 3).
For example, in type IIB theory the (scale of) 
$AdS_5 \times S^5$
solution is not modified  by the $R^4$ terms \rf{\bg}
only 
in the scheme  \gkt\ where  they have the form \jjj.

\newsec{$R^4$ terms in 11  dimensions}


Since the  invariant  ${\cal I}_2$  in \lopp\ contains
 the P-odd  CS type part $K$, 
 its  coefficient can not develop dilaton dependence
 without breaking
$B_2$   gauge invariance, i.e. its value  can not 
 be renormalized  from its 
 coupling-independent  one-loop  value \tse.
  Taking the limit $g_s \to \infty$
 this term   can then be 
 lifted to  a corresponding superinvariant in $D=11$ theory.
Assuming that the coefficient of the $J_0$ invariant \jo\ 
does  not receive  higher than one loop
 perturbative string corrections, 
it can be also lifted \refs{\gv,\rs,\kp,\ant}  to $D=11$ 
 (with  its tree-level part 
giving vanishing contribution). The resulting presence 
of the $t_8t_8R^4$ term in the M-theory effective action is 
indeed in agreement with  what follows directly  from the 
low-energy expansion of the 
4-graviton amplitude in $D=11$ supergravity \rf{\gree,\rs}.

In view of the above discussion, 
we  conclude that the effective action of the 
  $D=11$  theory should contain two distinct 
   $R^4$ superinvariants:
 (i)  $J_0$  with  $t_8t_8 R^4$  as its part, and (ii) 
${\cal I}_2$ which is a sum of the   $E_8$ and 
$\ep_{11} \C_3 R^4$ structures.
With this separation, 
the  coefficient in front of the $J_0$ term 
is then in agreement with
the 4-graviton amplitude (with the M-theory cutoff \gree),
and the coefficient of the ${\cal I}_2$ term (its 
$\C_3 R^4$ part)  is precisely  the one  implied 
by the M5 brane anomaly cancellation condition \duf.

Explicitly, 
the $D=11$ action  is then (cf. \tree,\loop)
$$
\S= \S_0 + \S_1 \ , 
$$
\eqn\ree{
\S_0 =
 {1\over 2\kappa^2_{11}} \int d^{11} x\ \sqrt {g} \  
\ \bigg[ R  - { 1 \ov 2\cdot 4!} F^2_4    - 
{ 1 \ov  6 \cd 3!
\cd (4!)^2 }  \ep_{11}  \C_3 F_4 F_4 \bigg] \ ,
\ \ 
}
\eqn\oop{
\S_1 =   b_1  T_2  \int d^{11} x \sqrt {g} \ 
\ \big ( J_0  -  2 {\cal I}_2 )  \ .  \ \ 
}
Here $F_{mnkl}  = 4 \del_{[m} \C_{nkl]}$ and
  the two  $R^4$ super-invariants are (see \inva,\gb,\comb) 
 \eqn\joe{
 J_0 = t_8\cdot t_8 RRRR + { 1\ov 4}  E_8  \ , \ \ \  \ \ \ \ 
 \   E_8=  {1\ov   3!} \ep_{11}\cdot \ep_{11} RRRR\ ,  
 }
\eqn\omb{
{\cal I}_2 = { 1\ov 4}  E_8  + 2 \ep_{11} \C_3 [\tr R^4 
 - { 1 \ov 4} (\tr  R^2)^2]  \ . }
 The constant $b_1=
  { 1 \ov (2\pi)^4 \cd 3^2 \cd 2^{13}}$ 
  is the same as in \nuu\ and the 
 10-d and 11-d parameters 
  are related as follows
  ($T_1$ and $T_2 $ are the string and the  membrane
  tensions)\foot{Note that  $B_2$ and $\C_3$ are canonically
  normalized, 
 so that 
 the 10-d invariant $ T_1 \int B_2 \wedge \tr ( \wedge R)^4 $
in \loop\  goes into  the 11-d  one \ 
$ T_2 \int \C_3 \wedge \tr ( \wedge R)^4 $, 
where in the  form notation
$B_2 = { 1 \ov 2} B_{mn} dx^m \wedge dx^n $, \ 
$\C_3 = { 1 \ov 3!} \C_{mnk} dx^m \wedge dx^n \wedge dx^k$,
$R^{ab} = { 1 \ov 2} R^{ab}_{mn} dx^m \wedge dx^n $.
Thus  $\S_1 $  \oop\ contains
 $  T_2  \int \C_3 \wedge \tr ( \wedge R)^4 $  
 with  the   coefficient 
 $  4\cd  3!\cd  2^4  b_1 =
  { 1 \ov  ( 2 \pi)^4\cd 3 \cd  2^6} $
 which is  the same as in \duf. }
 \eqn\rela{
   2\k^2_{11}  = (2\pi)^5 l^9_{11}\ ,\ \ \ \ 
   \kappa^2_{10} =   { \k^2_{11}  \ov  2\pi R_{11} }\ , \ \ \ 
  l_{11} = (2\pi g_s)^{1/3} \sqrt {\a'}\ , \ \ \ 
 R_{11}= g_s \sqrt {\a'}\ , }
 \eqn\mem {  
 T_2 = { 1 \ov  2\pi l_{11}^3  } = (2 \pi)^{2 /3}
  (2\k^2_{11})^{-1/3}\ , \ \ \ \  \ \ \ \ \ 
 T_1 = { 1 \ov 2\pi \a'} = 2 \pi R_{11} T_2  \ . }
 The  subleading $O(T_2)$ term \oop\ in the 
 effective action of 11-d  theory may contain also other 
 $O(R_{mn})$ and $O(F_4)$ terms.
The invariant $J_0$ 
(supplemented with appropriate $F_4$  dependent
terms)
may  be considered as a non-linear extension 
of the linearized ``$R^4$"  superinvariant 
in on-shell $D=11$ superspace \fer.
The P-even part of the  second   superinvariant starting with 
${\cal I}_2$ \omb\  may also include extra   
$O(F_4)$ terms.
Note that in the exterior  form notation ${\cal I}_2$ may be written as 
$$ {\cal I}_2 \ e^0 \we e^1 \we  ...\we e^{10}
\ = \ { 2 \ov 3} \ep_{11}
e \we e \we e 
\we R \we R \we R \we R  $$ 
\eqn\forma{
+  \ 
{ 2^5 \cd 3!  }\  \C_3 \we  \bigg[\tr (R\we R \we R \we R ) 
 - { 1 \ov 4} \tr  ( R\we R ) \we  \tr ( R \we R)  \bigg]
\ . } 
\newsec{$AdS_7 \times S^4$ solution and 
compactification on $S^4$}
The $D=11$ supergravity admits  the well known 
$AdS_7 \times S^4$ solution  with $F_4$ flux $N$ 
through $S^4$ \pilch. Compactifying on $S^4$, 
 one may derive 
the corresponding  $d=7$ supergravity  action,  
which gives the $O(N^3)$ contribution  \HS\ 
to  the conformal anomaly in the
corresponding 
boundary conformal (2,0) theory.

Let us  consider  how the presence  of the 
$R^4$ terms in the 11-d effective action   $\S_1$ \oop\ 
may  influence the existence of the 
$AdS_7 \times S^4$ solution  and expansion near it.
Using the on-shell  superspace 
 description  of 
11-d supergravity  and assuming that all local higher-order
corrections  to the  equations of motion 
can be written again in terms of the basic 
  on-shell supergravity superfield,
   it was argued in \kal\
that these  corrections cannot modify  the maximally
supersymmetric  $AdS_7 \times S^4$  solution. 
It  should be  possible  to see
 explicitly    that  adding the 
$J_0$ term in \oop\  (supplemented  with
  $F_4$ dependent terms as required by supersymmetry\foot{In
  addition
  to $F_4$ dependent terms (which may contain up to 8 powers 
  of $F_4$)  there are also $\del F_4$ dependent terms
  which  accompany  $t_8t_8R^4$ part of $J_0$
  in the 4-point S-matrix \dess\ (as suggested by
   the analysis 
  of tree-level 4-point scattering amplitudes in 11-d
  supergravity). These derivative  terms vanish on $AdS_7 \times S^4$
  background.}
  and chosen 
 in a  special 
  ``on-shell" scheme analogous but not equivalent\foot{Note
 that in contrast to $AdS_5 \times S^5$  space with equal radii
 the 11-d space 
  $AdS_7 \times S^4$ space with radii $1$ and $ \ha $
   is not conformally flat.} 
 to \jjj\ in 10-d theory)
 does not change  the leading-order 
 $AdS_7 \times S^4$ solution.
One  may view 
 $J_0$  as originating from a
restricted   superspace integral of 
$\W^4$, where $\W_{abcd}(x,\theta)$ is 
the on-shell supergravity  superfield 
\fer, which has the structure
$\W = F_4 + ... + \theta\theta ( \g...\g R + \g...\g F_4  F_4
+\g...\g  DF_4) + ...$\ 
($\g...\g$ stand for products of gamma matrices).
Then $J_0 \sim (R+ F_4F_4)^4$  and its first, second and third
variation over the metric evaluated on  $AdS_7 \times
S^4$ + $F_4$-flux
background ($R_{mn} \sim (F_4^2)_{mn}, \  \del F_4 =0$)
 will vanish,
essentially   as in  the case of $AdS_5 \times S^5$
solution of  type
  IIB theory   corrected by $J_0$
  term \bg\ (taken in the form \jjj).\foot{The vanishing of
 the first variation is equivalent to  the vanishing 
 of the  first correction to the 11-d supergravity 
 equations of motion $\g^{abc} D \W_{abcd} =0$ 
 due to the supercovariant constancy of $\W$ \kal.
 The argument of \kal\ should  certainly  apply  to the 
 first subleading  correction to the  11-d supergravity 
 equations of motion coming from $R^4$ terms in the
 action.}
 
The fact that the
$AdS_7 \times
S^4$ solution (and, in particular, the radii of its factors)
is not modified by the $J_0$ correction can be 
 also  represented as a consequence of the fact that 
 upon compactification 
 of the  11-d theory  on $S^4$ with  $F_4$ flux
  the $J_0$ term (taken in the special 
  ``on-shell" scheme) reduces to the Weyl tensor dependent
  $C^4$  term \jjj,   now defined  in 7 dimensions.\foot{The 
   tree + one-loop $J_0$ term 
  in type IIB theory leads to  the same 
   $C^4$ term \jjj\  in the 
   5-d effective  action  obtained by compactifying 
  the  type IIB theory   on $S^5$
  with $F_5$ flux.} 
  This term produces an $O(N)$ correction \gkt\ to the 
  leading $N^3$ term \KT\  in the entropy of 
  (2,0) theory describing
  multiple  M5 branes. 
  As in the 
 $AdS_5 \times S^5$  case in type IIB theory, 
  this  $C^4$ term does  not, however, 
   modify the expression for the
   conformal anomaly of the
  boundary conformal theory.\foot{It is important
    to stress for what follows 
  that in the above discussion 
   we treated $J_0$ \joe\ 
     as a whole, 
   without splitting it into $t_8t_8 R^4$ and $E_8$ parts. 
   It is  only that particular combination of $R^4$ terms 
    that takes the ``irreducible" 
    form \jjjk\  (cf. \myers), and thus should lead only 
    to $C^4 $ terms upon compactification 
    to $d=7$. At the same time, $E_8$ contains
    ``reducible" curvature contractions
     like $((R_{mnkl})^2)^2 + R (R_{mnkl})^3 + ...$
     and thus may, in principle, lead to 
     $O(R^n), \ n <4,$ terms upon compactification to $d=7$. }

  
Let us  now discuss the second invariant ${\cal I}_2$ \omb\ 
 in \oop.
It is easy to see that its P-odd  part $ \ep_{11} \C_3 [\tr R^4 
 - { 1 \ov 4} (\tr  R^2)^2]$ 
 does not modify the \adsss solution. Upon reduction on $S^4$ it
 leads to  $O(N)$ CS terms in $d=7$ action \harv.
 As for the $E_8$ part of  ${\cal I}_2$, 
 we shall assume that,   as  in the case of $J_0$, there
 exists an ``on-shell" scheme in which  this term, 
 supplemented with proper $F_4$-dependent terms,  
 also does not modify  the leading-order 
 \adsss solution.
 
  The  main point  is that upon compactification on $S^4$ 
   the $E_8$ term in \omb\
    should  produce 
   additional $R^3$  higher-derivative terms in the 7-d  effective 
   action 
   which, while not changing the  vacuum solution, 
  will give  subleading $O(N)$ 
corrections 
to the conformal  anomaly of the boundary CFT.\foot{These terms 
     will  give  also another $O(N)$  
      correction to the entropy 
  of (2,0) theory,  in addition 
   to the one coming from the  $J_0$  term \jjj\ found in \gkt.}


It   is  known that the  $\C_3R^4$  part of ${\cal I}_2$
\omb\  gives a subleading  $O(N)$
 correction 
 to the R-symmetry anomaly of the (2,0) theory \rf{\duf,\harv}.
 Since the R-symmetry and conformal anomalies 
should  belong to  the same 6-d supermultiplet,  
it is natural 
 to expect that the ``superpartner" of the $\C_3 R^4$ term, 
 i.e. the $E_8$ term  in ${\cal I}_2$, 
 should lead to an $O(N)$ correction
 to the conformal anomaly of the  boundary 6-d theory.
 This is  what  we are going to suggest  
 below.
  
  Since  we do not  know the  $F_4$
  (and $R_{mn}$) dependent
  terms which supplement $E_8$ to a superinvariant,  
to determine the terms in the 7-d action 
that originate from the $E_8$
part of the invariant $ {\cal I}_2$ in \oop\ 
 we shall use the following heuristic strategy.
 We shall start with $E_8$ and   compute it 
  in  the case when 
the 11-d space is a direct product,  $M^{11} = M^7 \times
M^4$. It is easy to see that 
\eqn\pro{
E_8 ( M^7 \times M^4) = 4 E_2 (M^4 ) E_6 (M^7)
 +  12 E_4 (M^4) E_4 (M^7)  \ , }
  where, as in \gb,  
  \eqn\yhu{ E_{2n} (M^d) \equiv  
  { 1 \ov (d-2n)!} \ep_{d}\cd \ep_{d} R^n \ , \ \ \ \ \ 
   \ \ \  d\geq 2n\ ,   }
  and $E_{2n} (M^d)
  =0
  $ for  $d < 2n$.
 In the case   when $M^4$ is a 4-sphere of radius $L$ 
 ($R_{S^4} = {12 \ov L^2}$)  and 
  $M^7$ has curvature $R$ 
 we get  
$$
E_8 ( M^7 \times S^4) =
 { 3 \cd 2^5 \ov L^2}   E_6 (M^7)  + 
 { 3^2 \cd 2^7 \ov L^4}   E_4 (M^7) $$  
\eqn\proo{ = \ { 3 \cd 2^5  \ov L^2}   \ep_7 \ep_7 RRR + 
 { 3 \cd 2^6  \ov L^4}   \ep_7 \ep_7 RR
  \ . }
  A remarkable property of the $E_8$ invariant is that 
  it does not produce a correction to the 
   cosmological or Einstein term in the 7-d
  action.
  
Next, we shall assume that when the same reduction 
is repeated 
for the analog of $E_8$ term   in a special  ``on-shell" 
scheme 
(i.e. for $E_8$  supplemented  by 
 $F_4$ and $R_{mn}$ dependent
 terms so that it does not produce a modification
 of  the leading-order \adsss solution)
then the  resulting terms in the
7-d action 
  will be the same as  in 
\proo\  but with  the curvature 
tensor  $R$ of $M^7$ replaced by its Weyl tensor $C$
part.

 
 In what follows we shall  consider  only on the 
 $E_6(M^7) \sim  C^3+...$  term  in \proo\ coming from $E_8$. 
 The reason is that we shall compute the corresponding contribution 
to the  scale anomaly  of the boundary theory  only modulo
$R_{mn}$-dependent terms, 
 but it is easy to see  that 
a   potential $C^2$ term 
   in the 7-d action  (coming from $E_4$ in \proo) 
can    lead  only to terms in the
 conformal anomaly which vanish
    when the 6-d boundary  space is Ricci flat. 
 
Choosing  the  normalization
in  which the radii of $AdS_7$ and $S^4$ are $1$
 and  $L=\ha$  so that 
  Vol$(S^4) = { 8 \pi^2 \ov 3} L^4
 ={ \pi^2\ov 6}$,   and  assuming that 
 the  value of the  quantized $F_4$ flux  is $N$, 
  we get   (see \rela,\mem\ and 
  \refs{\KT,\gkt})  
 \eqn\noot{
 { 1 \ov 2 \k_{11}^2} =   { N^3 \ov 2^8 \pi^5 L^9} = 
 { 2N^3 \ov \pi^5 }\ , \ \ \ \ \ 
 { 1 \ov 2 \k_{7}^2} = {  {\rm Vol}(S^4)\ov 2 \k_{11}^2} = 
  { N^3 \ov  3 \pi^3 }\ ,
 \ \ \ \ \ \  
 T_2 = { 2N\ov   \pi} \ . }
 The  relevant   $-\int [N^3 (R- 2\l) + N C^3 ]$ terms in the
   7-d  action\foot{Here we consider 
   the Euclidean signature and change overall 
   sign of the action, i.e. $\int R \to -\int R $.}
    are then
\eqn\sevv{ S^{(7)}  = 
 -{   N^3 \ov  3 \pi^3  } \int d^{7} x\ \sqrt {g} \  
\ ( R  +  30 ) 
   +   { \g  N \ov 3^2  \cd  2^{11}\cd  \pi^3 }   
   \int d^{7} x\ \sqrt {g}\   \hat E_6 + ...
\ ,  }
where  the explicit form of the $\hat E_6\sim C^3$ 
 correction term is 
(cf. \gb) 
 $$ 
\hat E_6 = (E_6)_{_{R_{mn}=0}} \ = \ 
\ep_7 \ep_7  C C C  $$
\eqn\cor{
= \ - 6!\delta^{n_1}_{[m_1}
\ldots \delta^{n_6}_{m_6]} C^{m_1 m_2}_{\ \ \ \ \ \ n_1 n_2}  
C^{m_3 m_4}_{\ \ \ \ \ \ n_3
n_4}C^{m_5 m_6}_{\ \ \ \ \ \ n_5 n_6} \ = \ -\  32\  (2I_1 +  I_2) \ ,   }
where $ I_1$ and $ I_2$ are defined  
as 
\eqn\few{
  I_1 = C_{amnb} C^{mpqn} C_p{}^{ab}{}_q \ ,\ \ \ \ 
   \ \ \ \ \ \
  I_2 =  C_{ab}{}^{mn}  C_{mn}{}^{pq} C_{pq}{}^{ab}
 \ .}
As follows from \proo\ the numerical coefficient 
$\g$ is 
\eqn\gag{
\g = 1 \ , }
but we shall  keep it arbitrary, given the 
uncertainties  in the above derivation of the correction term in 
\sevv\ (for example, 
the  presence of $(F_4)^2 (R_{mnkl})^3$  terms 
in $ {\cal I}_2$ would shift the value of $\g$).

\newsec{Conformal anomaly of (2,0) theory}
 
 Let us now determine the contribution of the $C^3$ correction  term
 in  the 7-d action \sevv\  which originated from the $E_8$ part of the 
 $\I_2$ superinvariant in the 11-d action \oop\
 to the conformal anomaly of the $d=6$  boundary 
 conformal  theory. We shall follow the same  method as used in
 \HS\ in computing the leading $N^3$ term in the
  anomaly.\foot{Similar computation 
  of subleading  corrections to conformal anomaly of 4-d 
 boundary  conformal field  theories 
 (with ${\cal N} <4$  supersymmetry) 
  coming from  $R^2$  curvature  terms in 
 5-d effective  action were  discussed in
 \refs{\blau,\imbim,\odin,\keh}.}
 We shall compute only the $O(N)$ contribution
 to the scale anomaly 
 (which is the same as integrated conformal anomaly,
  assuming topology of 6-space is
 trivial)
and   ignore terms 
 which  depend on $R_{mn}$,
  i.e. concentrate only on the  Weyl-invariant 
  non total derivative 
 $C^3$ terms (``type B" part) in the 6-d conformal anomaly.
 
 To obtain  the  conformal anomaly 
 one  is to  solve the 7-d equations for the
 metric (as in \noot\ we set the radius of $AdS_7$  to be 
 equal to 1)
 \eqn\meet{
 ds^2 = { 1 \ov 4} \r^{-2} d\r^2 + \r^{-1} g_{ij} (x,
 \r) dx^i dx^j \ , }
 evaluate the action on the solution
 $g=g_0(x) + \r g_2(x)  + ...$, 
 and compute its  variation  under the Weyl rescaling  of
 the 6-d boundary metric. The anomaly is essentially determined by 
 the  coefficient of the logarithmic divergence 
 produced by the integral over $\r$\  \HS. 
 In the present case of \sevv\ we find  (using \sevv,\cor\
 and $R_{AdS_7}  = -42$) 
   \eqn\sevvr{ S^{(7)}  = \int d^{6} x \bigg[ 
 {   N^3 \ov  3 \pi^3  }\cd 6\cd
  \int_\ep {d\r \ov \r^4}  \ \sqrt {g (x,\r)} \  
  -   \ { \g N \ov 3^2  \cd  2^{6}\cd  \pi^3 } \cd { 1 \ov 2} \cd 
    \int_\ep {d\r \ov  \r}
      \ \sqrt {g } \  (2I_1 + I_2)   + ... \bigg]
\ .  }
  Since \HS
    \eqn\gre{  6 \int_\ep {d\r \ov \r^4}\ \sqrt {g (x,\r)}\
 = \ \sqrt { g_0}\ \big[ a_0 (x) \ep^{-3} + ...
   - a_6 (x) \ln \ep\big]   + ... \ , } 
  the  anomaly is given by
 the sum of the $O(N^3)$ and $O(N)$ terms\foot{To obtain the 
 $O(N)$ contribution we  evaluate  the  $C^3$ term  
 in the 7-d action on the leading-order solution
 for the metric \meet\ (see \blau\ for a similar computation in
 the case of the $R^2_{mnkl}$ action in $d=5$), separate the 
 $C^3$ part depending on the 6-d metric $g_0$, and omit other 
 parts that depend on the Ricci tensor of $g_0$. }
  \eqn\ann{
 {\cal A}_{(2,0)} = {\cal A}^{N^3}_{(2,0)} + {\cal A}^{N}_{(2,0)}= 
 - {   N^3 \ov  3 \pi^3  }\cd  2 a_6  
 \ +  \ {\g  N \ov 3^2  \cd  2^{6}\cd  \pi^3 }
   \  (2I_1 + I_2  +  ...)  \ .}
   Here  $a_6 $   and $ I_1,I_2$  are evaluated for 
    the boundary
   metric $g_0$, and  
   dots stand for $O(N)$ \  $R_{mn}$-dependent 
   and total derivative 
   terms we are ignoring.
   
The result of \HS\ for the leading-order contribution
$ {\cal A}^{N^3}_{(2,0)}$  written 
as a sum of the type A (Euler), type B (Weyl invariant)
and scheme-dependent (covariant total derivative)
terms \rf{\bon,\schw}
is
\eqn\old{
 {\cal A}^{N^3}_{(2,0)}
=- {4N^3\ov (4\pi )^3\cd 3^2 \cd 2^5 }
\bigg[ E_{6}  + 8 
(12 I_{1}  +  3 I_{2}  - I_{3}) + O(\nabla_i J^i)  \bigg] \ ,
}
where
$E_6= \ep_6\ep_6 RRR$. The invariants  \  $I_1,I_2$ \few\ 
and $I_3$
\eqn\ferv{
 I_3 = C_{mnbc}   \nabla^2  C^{mnbc}
 + O(R_{mn})  + O(\nabla_i  J^i)
\ ,  }
  which form   the  basis of 3  Weyl invariants 
 are  the same  as  used in \bft.
 They are related 
  to the invariants used in \rf{\bon, \HS}
as follows:\  $E_{(6)},I_1,I_2$ and $I_3$ in \HS\ are 
equal to
\noindent
 ${ 1 \ov  3^3 \cd 2^{11}} E_6, -I_1,-I_2$ and
  $ -5 I'_3$, \ 
  $ I'_3= I_3 - { 8\ov 3  } ( 2 I_1 + I_2)  - { 1 \ov 12} E_6
  + O(\nabla_i J^i),$ 
  in terms of the invariants
  $E_6,I_1,I_2$ and $I_3$  used in \bft\ and  here.\foot{
  Our  curvature tensor $R^a_{\ bmn} = \del_m
\Gamma^a_{bn}-...$ has the opposite sign to
 that  of \HS. Note also that  
  \bft\ was   assuming  Euclidean
signature where $E_6$ is  defined as  $-\ep_6\ep_6 RRR$.}

We use this opportunity to point out  that the 
curvature 
invariant 
$I_3=-5 I'_3$ as defined in  \rf{\bon, \HS} is {\it  not}, 
in fact, covariant under Weyl transformations, contrary to what was assumed in 
\HS\  (this can be easily checked by 
computing it  for the   metric of a sphere $S^6$: 
one finds that  while  $I_1(S^6)= I_2(S^6)= 0$,
 \ \  $I'_{3}(S^6)\not=0$).
The proper  third  Weyl invariant of type $C  \nabla^2  C$
\ferv\ was  given   in \fef\ and is  equivalent 
 to the 
Weyl invariant $I_3$ used in \bft\ and here. 
Since $I_3$ of \HS\ or  $I_3'$ is a mixture of the true Weyl
invariants 
$I_1,I_2,I_3$ with $E_6$,
the separation of the leading  $N^3$    Weyl anomaly 
of the (2,0)  theory  \HS\ 
into  type A and type B parts  was not presented  correctly 
 in \HS. The correct separation was 
given in \bft\ and is used here.\foot{Note
that when $R_{mn}=0$
the two invariants -- $I_3'$ and $I_3$ -- 
 coincide, up to 
a covariant total derivative term. In fact, 
a separation of the conformal anomaly into type A and type B
parts becomes ambiguous on a Ricci flat background.}

Note that modulo  terms that vanish 
 for
$R_{mn}=0$  and total derivative terms,  
one has the following relations  (cf. \cor) 
\eqn\rell{E_6  = - \ 32\ ( 2I_1 +   I_2) +  O(R_{mn})
\ , \ \ \ \ \  \ \ \ 
I_3= 4 I_1 - I _2 +  O(R_{mn}) + O(\nabla_iJ^i)\  ,  }
so that ${\cal A}^{N^3}_{(2,0)}$ 
vanishes for $R_{mn}=0$,  as it should \HS.

Eq. \old\  is to be compared with the expression for the 
conformal anomaly  for  the  free  (2,0) tensor multiplet
found in \bft:
\eqn\teen{
 {\cal A}_{tens.}
=- {1\ov (4\pi )^3\cd 3^2 \cd 2^5 }
\bigg[ { 7 \ov 4} E_{6}  + 8 
(12 I_{1}  + 3 I_{2}  -  I_{3}) + O(\nabla_i J^i)  \bigg] \ .
}
As was concluded in \bft, the Weyl-invariant 
 (type B)  parts of  
 the  leading (2,0) theory anomaly  \old\  and 
the tensor multiplet anomaly \teen\  have 
exactly the same form,  
up to the overall  factor $4N^3$ in \old.

Since we have  found the $O(N)$ correction to the anomaly
of the (2,0) theory in \ann\  only modulo $R_{mn}$-dependent 
and total
derivative terms,  we  are   able to compare only 
 type B  anomalies, or  scale anomalies
(assuming  that the $d=6$  space has trivial topology,
 so that we can 
ignore  the integral of the Euler term $E_6$)
$$ {\bf A}_{(2,0)}=
\int d^6 x \sqrt {  g_0} \ {\cal
A}_{(2,0)}\ , \ \ \ \ \ 
{\bf A}_{tens.}=
\int d^6 x \sqrt {  g_0} \ {\cal
A}_{tens.}\ . $$
 Using  \rell\ 
to express $I_3$ in terms of $I_1$ and $I_2$, 
 we find from \old,\ann\ and \teen\ 
 \eqn\hyt{
{\bf A}^{N^3}_{(2,0)}=\ - {4N^3\ov (4\pi )^3\cd 3^2}
 \int d^6 x \sqrt {  g_0}
\  (  2 I_1 + I_2 )  \ ,
}
\eqn\hfyt{
{\bf A}^N_{(2,0)}= 
 \ { \g N \ov (4\pi )^3\cd 3^2 }  \int d^6 x \sqrt {  g_0}
\ ( 2 I_1 + I_2) \ , 
}
 \eqn\hyt{
{\bf A}_{tens.}
=\ -{1\ov (4\pi )^3\cd 3^2  } \int d^6 x \sqrt {  g_0}
\  (  2 I_1 + I_2 )  \ .
}
The total scale anomaly of the (2,0) theory 
following from \sevv,\ann\ is then 
\eqn\hyit{
{\bf A}_{(2,0)}=   {\bf A}^{N^3}_{(2,0)}  
+ {\bf A}^N_{(2,0)}   =
\ - {4N^3 - \g N \ov (4\pi )^3\cd 3^2} \int d^6 x
\sqrt {  g_0}\  (  2 I_1 + I_2 )  \ .  
}
Equivalently, 
\eqn\yit{
{\bf A}_{(2,0)}  =
\ - {4(N^3 - N) \ov (4\pi )^3\cd 3^3} \int d^6 x
\sqrt {  g_0}\  (  2 I_1 + I_2 ) \  + \ (4-\g) 
 N\  {\bf A}_{tens.} \ . 
}
Thus  if the true value of $\g$  is 
3  instead of the  naive  value $1$ \gag\
which follows directly  from  reduction of $E_8$  \proo, 
ignoring possible $F_4$-dependent ($F^2_4 R^3$) terms 
in the 11-d  super-invariant  ${\cal I}_2$,  
then ${\bf A}_{(2,0)}$ 
reproduces
the scale anomaly \hyt\ of  a single (2,0) tensor multiplet.
  This $N=1$  relation  should 
be   expected, given that a similar correspondence 
is true for the R-symmetry anomalies \harv\ (see below).  
 Though we are unable to show that $\g=3$ does follow
 from the $d=7$  reduction of the  11-d 
 super-invariant  ${\cal I}_2$ containing  P-odd 
 $\C_3 R^4$ term,
 we find it    remarkable that the required
  value of $\g$ differs from the naive value 1 
simply by factor of 3.\foot{In the original 
version of the present paper we mistakenly 
 used  the basis 
of type B invariants  including $I_3$ of \HS\ instead 
of the correct invariant of \bft\ and as a result 
got the $O(N)$ term with extra coefficient 3,
concluding that $\g=1$ gives already the desired 
coefficient $4N^3-3N$ in \hyit.}  \foot{Note that if we were comparing  the full local  
conformal anomalies  evaluated  for  $R_{mn}=0$ 
then, since  the $N^3$ contribution \old\ vanishes in this case, 
we  would need $\g= {3 \ov 4}$ in order to reproduce the non-zero
$R_{mn}=0$ value of the tensor multiplet anomaly \teen\
by the $N=1$ limit of the $O(N)$ term in \ann.}


Making a natural  conjecture
 that the same  relation  
 ${\cal A}_{tens.} = ({\cal A}_{(2,0)})_{N=1}$ should be
true  between the full expressions for the conformal anomalies 
of the (2,0) theory and tensor multiplet, one  
 can make a  prediction about 
  the complete structure of the  $O(N) $ term in 
 the  (2,0) theory anomaly ${\cal A}_{(2,0)}$ \ann\ 
 (cf. \old,\teen)\foot{The shift of the coefficient of the $E_6$ term
 in the conformal anomaly
 seems to imply a contradiction  between our assumption 
 that the $R^4$  terms  in the 11-d action \oop\ 
 do not change the scale of \adsss solution
 (i.e. that the value of the 7-d action \sevv\ evaluated 
 on the $AdS_7$ solution is not changed), and the claim 
 of \imbim\ that the coefficient of the type A (Euler) term 
 in the anomaly of a generic effective theory 
 is  determined only by the value of the action on the 
 $AdS$ solution.}
  \eqn\tepn{
 {\cal A}_{(2,0)}
=- {1\ov (4\pi )^3\cd 3^2 \cd 2^5 }
\bigg[ (4 N^3 -  { 9 \ov 4} N) E_{6}  
+ (4 N^3 - 3 N) \cd  8  (12 I_{1}  + 3 I_{2}  - I_{3}) 
+ O(\nabla_i J^i)  \bigg] \ , 
}
or, equivalently, 
\eqn\uepn{
 {\cal A}_{(2,0)}
=- {N^3-N \ov (4\pi )^3\cd 3^2 \cd 2^3 }
\bigg[  E_{6}  
+  8  (12 I_{1}  +  3I_{2}  -  I_{3} ) 
 +  O(\nabla_i J^i)  \bigg] 
 +   N  {\cal A}_{tens.} \ .
}
Using \rell, we can rewrite   \tepn\ also as 
 \eqn\oepn{
 {\cal A}_{(2,0)}
=- {N\ov (4\pi )^3\cd 3 \cd 2^7 }
\bigg[   E_{6}  + O(R_{mn})  + O(\nabla_i J^i)  \bigg] \ , 
}
in agreement with the fact that for $R_{mn}=0$ 
the conformal anomaly of the tensor multiplet becomes
\refs{\bft,\ft}\ 
$ {\cal A}_{tens.}
=- {1\ov (4\pi )^3\cd 3 \cd 2^7 }
\big[ E_{6}  + O(\nabla_i J^i)  \big]$. 

 

It is useful to compare the above expressions \hyit,\tepn\
with the previously known results for the R-symmetry anomalies of the 
interacting (2,0) theory and free tensor multiplet theory.
The 1-loop  effective action $\G$  for  a free 6-d tensor multiplet 
in a background of 6-d  Lorentz 
curvature $R$
 and $SO(5)$ R-symmetry gauge field  $F$ 
 has local $SO(6)$ and $SO(5)$ anomalies.
 They 
 satisfy the  descent relations 
  $d ( \delta \G)= \delta I_7, \
 \ I_8 = d I_7$,  with  the 8-form anomaly polynomial $I_8$ being  
 \rf{\alv,\witte}
 \eqn\ank{
 I_8^{tens.} (F,R)
 = {1 \ov 3 \cd 2^4} \bigg[ p_2 (F) - p_2(R) + { 1 \ov 4} 
 [ p_1(F)- p_1(R) ]^2 \bigg] \ , }
 with  (here $F^2\equiv F \wedge F$, etc.) 
 \eqn\uank{
p_1 (F) = { 1 \ov 2}\   \tr\  \bar F^2  \ , 
\ \ \ \ \ \ \ \  
p_2 (F) = - { 1 \ov 4} 
  \bigg( \tr\ \bar  F^4 -
   { 1 \ov 2}  \tr\ \bar F^2\wedge  \tr\ \bar F^2 \bigg)\ , 
   \ \ \ \  \bar F =  { i \ov 2 \pi}\  F \ . 
}
The corresponding  anomalies  of the  interacting (2,0)  theory 
describing multiple M5 branes  derived  
(by assuming that the total M5-brane anomaly
 + inflow anomaly should
cancel)
from the 11-d supergravity action \ree\ with  the 
$R^4$ correction term  \oop\ 
is 
\rf{\harv } 
\eqn\ank{
 I_8^{(2,0)} (F,R) =
 { 1 \ov 3 \cd 2^4}\bigg[\  (2N^3 - N) \  p_2(F)\ 
 - \   N  \ p_2(R) \ + \ { 1 \ov 4}  N \
 [ p_1(F)- p_1(R) ]^2 \bigg] \ . }
Here the $O(N^3)$ term comes \freed\ from the CS term 
in supergravity action  \ree\  and the $O(N)$ term 
\rf{\duf,\witte}-- from the  P-odd  $\C_3 R^4$ part of the 
 superinvariant ${\cal I}_2$ \oop,\omb. 
Equivalently, 
\eqn\ank{
 I_8^{(2,0)} (F,R) =
{ 1 \ov 3 \cd 2^3} (N^3 - N) \ p_2(F)\ \ + \ \ N \ I_8^{tens.} (F,R) 
 \ . }
Thus for $N=1$ 
 the anomaly of the (2,0) theory is the same as the anomaly of 
 a single tensor multiplet.
 This is the same  type of   a relation we have established above 
 (cf.  \yit) 
 for the scale anomalies, with the crucial $O(N)$ 
 contribution coming from the P-even $E_8$ part  of the 
 superinvariant ${\cal I}_2$ \omb.
  This is  obviously 
  consistent with the fact 
   that R-symmetry and
  conformal anomalies
  should be  parts of  the same 6-d supermultiplet.

\bigskip
\noindent {\bf Acknowledgements}
\medskip
We are grateful to S. Frolov  for a collaboration 
at an initial
stage and many useful  discussions. 
 We would like also to  acknowledge  J. Harvey, 
  P. Howe, K. Intriligator, R. Metsaev, Yu. Obukhov,
  H. Osborn,   T. Petkou   and M. Shifman  for
helpful discussions and comments.
This work was  supported in part by
the DOE grant DOE/ER/01545, 
 EC TMR grant ERBFMRX-CT96-0045, 
INTAS grant No. 99-0590,  NATO grant PST.CLG 974965
and PPARC SPG grant  PPA/G/S/1998/00613.

\vfill\eject
\listrefs
\end